\documentclass[a4paper,UKenglish]{lipics-v2016}
\pdfoutput=1
\usepackage{microtype}%
\usepackage{amssymb}
\usepackage{longtable}
\usepackage{amsmath}
\usepackage{algorithm}
\usepackage{algpseudocode}
\newfloat{algorithm}{t}{lop}
\usepackage{etoolbox}
\usepackage{verbatim}
\usepackage{soul}

\renewcommand{\epsilon}{\varepsilon}

\newcommand{\floor}[1]{\left\lfloor #1 \right\rfloor}

\newcommand{\prob}{\ensuremath{\mathsf{Pr}}}
\newcommand{\expect}{\ensuremath{\mathsf{E}}}

\newcommand{\BoundedSAT}{\ensuremath{\mathsf{BSAT}}}

\newcommand{\var}{\ensuremath{\mathsf{V}}}

\newcommand{\SharpP}{\ensuremath{\mathsf{\#P}}}

\newcommand{\SAT}{\ensuremath{\mathsf{SAT}}}

\newcommand{\CheckSAT}{\ensuremath{\mathsf{CheckSAT}}}

\newcommand{\ApproxMC}{\ensuremath{\mathsf{ApproxMC}}}

\newcommand{\ScalApproxMC}{\ensuremath{\mathsf{ApproxMC2}}}
\newcommand{\hiThresh}{\ensuremath{\mathsf{hiThresh}}}
\newcommand{\cellCount}{\ensuremath{\mathsf{cellCount}}}
\newcommand{\emptyList}{\ensuremath{\mathsf{emptyList}}}
\newcommand{\iter}{\ensuremath{\mathsf{iter}}}
\newcommand{\ScalApproxMCCore}{\ensuremath{\mathsf{ApproxMC2Core}}}
\newcommand{\AddToList}{\ensuremath{\mathsf{AddToList}}}
\newcommand{\solCount}{\ensuremath{\mathsf{solCount}}}
\newcommand{\FindMedian}{\ensuremath{\mathsf{FindMedian}}}

\newcommand{\FibBinSearch}{\ensuremath{\mathsf{LogSATSearch}}}
\newcommand{\lowerFib}{\ensuremath{\mathsf{lowerFib}}}
\newcommand{\upperFib}{\ensuremath{\mathsf{upperFib}}}
\newcommand{\FailRecord}{\ensuremath{\mathsf{FailRecord}}}

\newcommand{\SymbolicDNFApproxMC}{\ensuremath{\mathsf{SymbolicDNFApproxMC}}}
\newcommand{\SymbolicDNFApproxMCCore}{\ensuremath{\mathsf{SymbolicDNFApproxMCCore}}}

\newcommand{\enumNext}{\ensuremath{\mathsf{enumNextREX}}}
\newcommand{\Reduce}{\ensuremath{\mathsf{Extract}}}

\newcommand{\nextGrayBit}{nextGrayBit}
\newcommand{\SampleBase}{\ensuremath{\mathsf{SampleBase}}}

\newcommand{\rfha}{\Omega_{\vect{y}}}%
\newcommand{\rf}{\set{R_{\phi}}}

\newcommand{\minClsWidth}{\ensuremath{\mathsf{w}}}
\newcommand{\RRE}{Reduced Row-Echelon }

\newcommand{\numVars}{\ensuremath{\mathsf{n}}}
\newcommand{\numCubes}{\ensuremath{\mathsf{m}}}
\newcommand{\numHashVars}{\ensuremath{\mathsf{q}}}

\newcommand{\numHashConstraints}{\ensuremath{\mathsf{p}}}
\newcommand{\ha}{\ensuremath{\mathsf{\vect{z}}}}

\newcommand{\cs}{\ensuremath{\mathsf{c_{\vect{x}}}}}
\newcommand{\sL}{low}
\newcommand{\sH}{hi}
\newcommand{\sI}{sI}
\newcommand{\f}{f}
\newcommand{\voplus}{\oplus}
\newcommand{\cube}[1]{\phi^{C#1}}

\newcommand{\width}[1]{{width[\cube{#1}]}}
\newcommand\vect[2][]{%
	\ifstrempty{#1}{%
		\boldsymbol{#2}
	}{%
	\boldsymbol{#1}^{[#2]}
}%
}
\newcommand{\mat}[2][]{%
	\ifstrempty{#1}{%
		\boldsymbol{#2}
	}{%
	\boldsymbol{#1}^{[#2]}
}%
}
\newcommand{\set}[1]{#1}
\newcommand{\D}{\ensuremath{\mathsf{D}}}
\newcommand{\G}{\ensuremath{\mathsf{G}}}
\newcommand{\M}{\ensuremath{\mathsf{E}}}

\algnotext{EndFor}
\algnotext{EndIf}
\algnotext{EndWhile}

\title{On Hashing-Based Approaches to Approximate DNF-Counting\footnote{The author list has been sorted alphabetically by last name; this should not be used to determine the extent of authors’ contributions}}
\titlerunning{On Hashing-Based Approaches to Approximate DNF-Counting} %

\author[1]{Kuldeep S. Meel}
\author[2]{Aditya A. Shrotri}
\author[3]{Moshe Y. Vardi}
\affil[1]{National University of Singapore, Singapore, Singapore\\
  \texttt{meel@comp.nus.edu.sg}}
\affil[2]{Rice University, Houston, USA\\
  \texttt{as128@rice.edu}}
\affil[3]{Rice University, Houston, USA\\
  \texttt{vardi@rice.edu}}
\authorrunning{K.\,S. Meel, A.\,A. Shrotri and M.\,Y. Vardi} %

\Copyright{Kuldeep S. Meel, Aditya A. Shrotri and Moshe Y. Vardi}%


\begin{document}

\maketitle

\begin{abstract}
  Propositional model counting is a fundamental problem in artificial 
  intelligence with a wide variety of applications, such as probabilistic 
  inference, decision making under uncertainty, and probabilistic databases.
  Consequently, the problem is of theoretical as well as practical interest. 
  When the constraints are expressed as DNF formulas, Monte Carlo-based techniques
  have been shown to provide a fully polynomial randomized approximation 
  scheme (FPRAS). For CNF constraints, hashing-based approximation techniques
  have been demonstrated to be highly successful. Furthermore, it was shown 
  that hashing-based techniques also yield an FPRAS for DNF counting without 
  usage of Monte Carlo sampling. Our analysis, however, shows that the 
  proposed hashing-based approach to DNF counting provides poor time 
  complexity compared to the Monte Carlo-based DNF counting techniques. Given
  the success of hashing-based techniques for CNF constraints, it is natural 
  to ask: Can hashing-based techniques provide an efficient FPRAS for DNF 
  counting? In this paper, we provide a positive answer to this question. 
  To this end, we introduce two novel algorithmic techniques: \emph{Symbolic Hashing} and \emph{Stochastic Cell Counting}, along with a new hash family of \emph{Row-Echelon hash functions}. These innovations allow us to design a hashing-based FPRAS for DNF counting 
  of similar complexity (up to polylog factors) as that of prior works. Furthermore, we expect 
  these techniques to have potential applications beyond DNF counting.
\end{abstract}

\section{Introduction}
\label{sec: intr} 
Propositional model counting is a fundamental problem in artificial intelligence with a wide range of applications including probabilistic inference, databases, decision making under uncertainty, and the like~\cite{Bacchus2003,dalvi2007efficient,DH07,SangBeamKautz2005}. Given a Boolean formula $\phi$, the problem of propositional model counting , also referred to as \#SAT, is to compute the number of solutions of $\phi$~\cite{Valiant79}. Depending on whether $\phi$ is expressed as a CNF or DNF formula, the corresponding model counting problems are denoted as \#CNF or \#DNF, respectively. Both \#CNF and \#DNF have a wide variety of applications. For example, probabilistic-inference queries reduce to solving \#CNF instances~\cite{Bacchus2003,PD06,Roth1996,SangBeamKautz2005}, while evaluation of queries for probabilistic database reduce to \#DNF instances~\cite{dalvi2007efficient}. Consequently, both \#CNF and \#DNF have been of theoretical as well as practical interest over the years~\cite{Jerr,KarpLuby1989,Sang04combiningcomponent,Stockmeyer83}. In his seminal paper, Valiant~\cite{Valiant79} showed that both \#CNF and \#DNF  are \#P-complete, a class of problems that are believed to be intractable in general. 

Given the intractability of \#CNF and \#DNF, much of the interest lies in the approximate variants of \#CNF and \#DNF, wherein for given tolerance and confidence parameters $\varepsilon$ and $\delta$, the goal is to compute an estimate $C$ such that $C$ is within a $(1+\varepsilon)$ multiplicative factor of the true count with confidence at least $1-\delta$. While both \#CNF and \#DNF are \#P-complete in their exact forms, the approximate variants differ in complexity: approximating \#DNF can be accomplished in fully polynomial randomized time~\cite{dagum2000optimal,KarpLuby1983,KarpLuby1989}, but approximate \#CNF is NP-hard~\cite{Stockmeyer83}. Consequently, different techniques have emerged to design scalable approximation techniques for \#DNF and \#CNF.

In the context of \#DNF, the works of Karp, Luby, and Madras~\cite{KarpLuby1983,KarpLuby1989} led to the development of highly efficient Monte-Carlo based techniques, whose time complexity is  linear in the size of the formula. On the other hand, hashing-based techniques have emerged as a scalable approach to the approximate model counting of CNF formulas~\cite{CMV13b,CMV16,EGSS13c,GSS06,Stockmeyer83}, and are effective even for problems with existing FPRAS such as \emph{network reliability}~\cite{duenas2017counting}. These hashing-based techniques employ 2-universal hash functions to partition the space of satisfying solutions of a CNF formula into cells such that a randomly chosen cell contains only a {\emph small} number of solutions. Furthermore, it is shown that the number of solutions across the cells is {\em roughly equal} and, therefore, an estimate of the total count can be obtained by counting the number of solutions in a cell and scaling the obtained count by the number of cells. Since the problem of counting the number of solutions in a cell when the number of solutions is {\em small} can be accomplished efficiently by invoking a {\SAT} solver, the hashing-based techniques can take advantage of the recent progress in the development of efficient {\SAT} solvers. Consequently, algorithms such as {\ApproxMC}~\cite{CMV13b,CMV16} have been shown to scale to instances with hundreds of thousands of variables. 

While Monte Carlo techniques introduced in the works of Karp et al.~have shown to not be applicable in the context of approximate \#CNF~\cite{KarpLuby1989}, it was not known whether hashing-based techniques could be employed to obtain efficient algorithms for \#DNF.   Recently, significant progress in this direction was achieved by  Chakraborty, Meel and Vardi~\cite{CMV16}, who showed that hashing-based framework of {\ApproxMC} could be employed to obtain FPRAS for \#DNF counting%
\footnote{It is worth noting that several hashing-based algorithms based on~\cite{EGSS13c,trevisan2002lecture} do not lead to FPRAS schemes for \#DNF despite close similarity to Chakraborty et al.'s approach}. There is, however, no precise complexity analysis in~\cite{CMV16}. In this paper, we provide a complexity analysis of the proposed scheme of Chakraborty et al., which is worse than quartic in the size of formula. In comparison, state-of- the-art approaches achieve complexity linear in the number of variables and cubes for \#DNF counting. This begs the question: {\em How powerful is the hashing-based framework in the context of DNF counting? In particular, can it lead to algorithms competitive in runtime complexity with state-of-the-art? }

In this paper, we provide a positive answer to this question. To achieve such a significant reduction in complexity, we offer three novel algorithmic techniques: (i) A new class of 2-universal hash functions that enable fast enumeration of solutions using Gray Codes, (ii) Symbolic Hashing, and (iii)  Stochastic Cell Counting. These techniques allow us to achieve the complexity of $ \tilde{O}(\numCubes\numVars\log(1/\delta)/\epsilon^{2}) $, which is within polylog factors of the complexity achieved by Karp et al.~\cite{KarpLuby1989}. Here, $ \numCubes $ and $ \numVars $ are the number of cubes and variables respectively while $ \epsilon $ and $ \delta $ are the tolerance and confidence of approximation. Furthermore, we believe that these techniques are not restricted to \#DNF. Given recent breakthroughs achieved in the development of hashing-based CNF-counting techniques, we believe our techniques have the potential for a wide variety of applications. 

The rest of the paper is organized as follows: we introduce notation in section \ref{sec:prelim} and discuss related work in section \ref{sec:rel}. We describe our main contributions in section \ref{sec:body}, analyze the resulting algorithm in section \ref{sec:ana} and discuss future work and conclude in section \ref{sec:conc}.

\section{Preliminaries}	
\label{sec:prelim}
	\subsection*{DNF Formulas and Counting}
	We use Greek letters $ \phi $, $ \theta $ and $ \psi $ to denote boolean formulas. A formula $ \phi $ over boolean variables $ x_{1}, x_{2},\dots, x_{\numVars} $ is in Disjunctive Normal Form (DNF) if it is a disjunction over conjunctions of variables or their negations. We use $ \set{X} $ to denote the set of variables appearing in the formula. Each occurrence of a variable or its negation is called a \emph{literal}. Disjuncts in the formula are called \emph{cubes} and we denote the $ i^{th} $ cube by $ \cube{i} $. Thus $ \phi = \cube{1} \lor \cube{2} \lor ... \lor \cube{\numCubes}$ where each $ \cube{i} $ is a conjunction of \emph{literals}. We will use $ \numVars $ and $ \numCubes $ to denote the number of variables and number of cubes in the input DNF formula, respectively. The number of literals in a cube $ \cube{i} $ is called its \emph{width} and is denoted by $\width{i}$. 

	An assignment to all the variables can be represented by a vector $ \vect{x} \in \{0,1\}^{\numVars}$ with $ 1 $ corresponding to $true$ and $ 0 $ to $false$. $ \set{U} = \{0,1\}^{\numVars}$ is the set of all possible assignments, which we refer to as the \emph{universe} or state space interchangeably. An assignment $ \vect{x} $ is called a satisfying assignment for a formula $ \phi $ if $ \phi $ evaluates to $true$ under $ \vect{x} $. In other words $\vect{x}$ satisfies $ \phi $ and is denoted as $ \vect{x} \models \phi $. Note that an assignment $ \vect{x} $ will satisfy a DNF formula $ \phi $ if $ \vect{x} \models \cube{i} $ for some $ i $. The DNF-Counting Problem is to count the number of satisfying assignments of a DNF formula.
	
	Next, we formalize the concept of a counting problem. Let $ \set{R} \subseteq \{0,1\}^{*} \times \{0,1\}^{*} $ be a relation which is decidable in polynomial time and there is a polynomial $p$ such that for every $ (s,t) \in R $ we have $ |t| \le p(|s|)$. The decision problem corresponding to $ R $ asks if for a given $ s $ there exists a $ t $ such that $ (s,t) \in \set{R} $. Such a problem is in NP. Here, $ s $  is a called the \emph{problem instance} and $ t $ is called the \emph{witness}. We denote the set of all witnesses for a given $ s $ by $ \set{R_{s}} $. The \emph{counting problem} corresponding to $ \set{R} $ is to calculate the size of the witness set $ |\set{R_{s}}| $ for a given $ s $. Such a problem is in \SharpP \cite{Valiant79}.
	The DNF-Counting problem is an example of this formalism:  A formula $ \phi $ is a problem instance and a satisfying assignment $ \vect{x} $  is a witness of $ \phi $. The set of satisfying assignments or the solution space is denoted $ \set{R_{\phi}} $ and the goal is to compute $ |\rf| $. It is known that the problem is \SharpP-Complete, which is believed to be intractable~\cite{Toda89}. Therefore, we look at what it means to efficiently and accurately approximate this problem.
	
	A \emph{fully polynomial randomized approximation scheme} (FPRAS) is a randomized algorithm that takes as input a  problem instance $ s $, a tolerance $ \epsilon \in (0,1) $ and confidence parameter $ \delta \in (0,1) $ and outputs a random variable $ c $ such that 
	$\Pr[\frac{1}{1+\epsilon}|\set{R_{s}}| \le c \le (1+\epsilon)|\set{R_{s}}|] \ge 1 - \delta
	$
	and the running time of the algorithm is polynomial in $ |s| $, $ 1/\epsilon $, $ \log (1/\delta) $ \cite{KarpLuby1983}. 
	Notably, while exact DNF-counting is inter-reducible with exact CNF-counting, the approximate versions of the two problems are not because multiplicative approximation is not closed under complementation.
	\subsection*{Matrix Notation}
	We use $ x $,$ y,z,\dots $ to denote scalar variables. We use subscripts $ x_{1},x_{2},\dots $ as required. In this paper we are dealing with operations over the boolean ring, where the variables are boolean, 'addition' is the XOR operation ($ \oplus $) and 'multiplication' is the AND operation ($ \land $). We use the letters $ i $,$ j $,$ k,l $ as indices or to denote positions. We denote sets by non-boldface capital letters. We use capital boldface letters $ \mat{A} $,$ \mat{B},\dots $ to denote matrices, small boldface letters $ \vect{u} $, $ \vect{v} $, $ \vect{w}, \dots$ to denote vectors. $ \mat[A]{\numHashConstraints\times \numHashVars} $ denotes a matrix of $ \numHashConstraints $ rows and $ \numHashVars $ columns, while $ \vect[u]{\numHashVars} $ denotes a vector of length $ \numHashVars $. $ \vect[0]{\numHashVars} $ and $ \vect[1]{\numHashVars} $ are the all 0s and all 1s vectors of length $n$, respectively. We omit the dimensions when clear from context. $ \vect{x}[i] $ denotes the $i^{th}$ element of $\vect{x}$, while $ \mat{A}[i,j] $ denotes the element in the $i$th row and $j^{th}$ column of $ \mat{A} $. $ \mat{A}[r_{1}:r_{2},c_{1}:c_{2}] $ denotes the sub-matrix of $ \mat{A} $ between rows $ r_{1} $ and $ r_{2} $ excluding $ r_{2} $ and columns $ c_{1} $ and $ c_{2} $ excluding $ c_{2} $. Similarly $ \vect{v}[i:j] $ denotes the sub-vector of $ \vect{v} $ between index $ i $ and index $ j $ excluding $ j $.  The $ i^{th} $ row of $ \mat{A} $ is denoted $ \mat{A}[i,:] $ and $ j^{th} $ column as $ \mat{A}[:,j] $. The $ \numHashConstraints \times (\numHashVars_{1}+\numHashVars_{2}) $ matrix formed by concatenating rows of matrices $ \mat[A]{\numHashConstraints \times \numHashVars_{1}} $ and $ \mat[B]{\numHashConstraints \times \numHashVars_{2}} $ is written in block notation as $ [\mat{A} \,\,|\,\, \mat{B}] $, while $ [\frac{\mat{A}}{\mat{B}}] $ represents concatenation of columns. Similarly the $ (\numHashVars_{1}+\numHashVars_{2})$-length concatenation of vectors $\vect[v]{\numHashVars_{1}}$ and $\vect[w]{\numHashVars_{2}}$ is $ [\vect{v} \,\,|\,\, \vect{w}] $. The dot product between matrix $ \mat{A} $ and vector $ \vect{x} $ is written as $ \mat{A}.\vect{x} $. The vector formed by element-wise XOR of vectors $ \vect{v} $ and $ \vect{w} $ is denoted $ \vect{v} \voplus \vect{w} $. 	
	
	\subsection*{Hash Functions}
	A hash function $ h: \{0,1\}^{\numHashVars} \rightarrow \{0,1\}^{\numHashConstraints} $ partitions the elements of of the domain $\{0,1\}^{\numHashVars}$ into $ 2^{\numHashConstraints} $ cells. $ h(\vect{x}) = \vect{y} $ implies that $ h $ maps the assignment $ \vect{x} $ to the cell $ \vect{y} $. $ \set{h^{-1}(\vect{y})} = \{\vect{x} | h(\vect{x}) = \vect{y}\} $ is the set of assignments that map to the cell $ \vect{y} $. In the context of counting, 2-universal families of hash functions, denoted by $ H(\numHashVars, \numHashConstraints, 2)$, are of particular importance. When $h$ is sampled uniformly at random from $ H(\numHashVars, \numHashConstraints, 2) $, 2-universality entails
	\begin{enumerate}
		\item $ \Pr[h(\vect{x}_{1}) = h(\vect{x}_{2})] \le 2^{-\numHashConstraints} $ for all $ \vect{x}_{1} \ne \vect{x}_{2} $
		\item $ \Pr[h(\vect{x}) = \vect{y}] = 2^{-\numHashConstraints} $ for every $ \vect{x} \in \{0,1\}^{\numHashVars} $ and $ \vect{y} \in \{0,1\}^{\numHashConstraints} $.
	\end{enumerate} 

	Of particular interest is the random XOR family of hash functions, which is defined as $H_{XOR}(\numHashVars,\numHashConstraints)=\{\mat{A}.\vect{x}\voplus\vect{b} \,\, | \,\,  \mat{A}[i,j] \in \{0,1\} \,\,and \,\, \vect{b}[i] \in \{0,1\} \,\, ;\,\, 0\le i < \numHashConstraints,\,\, 0\le j < \numHashVars\}$.  Selecting $ \mat{A}[i,j]s $ and $ \vect{b}[i] $s randomly from $ \{0,1\} $ is equivalent to drawing uniformly at random from this family. A pair $\mat{A} $ and $\vect{b}$ now defines a hash function $h_{\mat{A},\vect{b}}$ as follows: $h_{\mat{A},\vect{b}}(x)= \mat{A}.\vect{x} \voplus \vect{b}$.  This family was shown to be 2-universal in \cite{carter1977universal}.  For a hash function $h\in H_{XOR}(\numHashVars,\numHashConstraints)$, we have that $ h(\vect{x}) = \vect{y} $ is a system of linear equations modulo 2: $ \mat{A}.\vect{x} \voplus \vect{b} = \vect{y} $.  From another perspective, it can be viewed as a boolean formula $ \psi = \bigwedge_{i=1}^{\numHashConstraints} (\bigoplus_{j=1}^{\numHashVars} (\mat{A}[i,j]\land \vect{x}[j])) \oplus \vect{b}[i] = \vect{y}[i] $. The solutions to this formula are exactly the elements of the set $ \set{h^{-1}(\vect{y})} $. 
	
	\subsection*{Gaussian Elimination}
	 Solving a system of linear equations over $ \numHashVars $ variables and $ \numHashConstraints $ constraints can be done by row reduction technique known variously as \emph{Gaussian Elimination} or \emph{Gauss-Jordan Elimination}. A matrix is in \emph{Row-Echelon form} if rows with at least one nonzero element are above any rows of all zeros.  The matrix is in \emph{Reduced Row-Echelon form} if, in addition, every leading non-zero element in a row is 1 and is the only nonzero entry in its column.  We refer to the technique for obtaining the \emph{Reduced Row-Echelon} form of a matrix as Gaussian Elimination. We refer the reader to any standard text on linear algebra (cf., \cite{strang1993introduction}) for details. For a matrix in \RRE form, the row-rank is simply the number of non-zero rows. 

	For a system of linear equations $ \mat{A}.\vect{x} \voplus \vect{b}= \vect{y} $, if the row-rank of the augmented matrix is same as row-rank of $ \mat{A} $, then the system is consistent and the number of solutions is $ 2^{\numHashVars - rowrank(\mat{A})} $ where $ \numHashVars $ is the number of variables in the system of equations. 
	Moreover, if $ \mat{A} $ is in \RRE form, then the values of the variables corresponding to leading $ 1 $s in each row are completely determined by the values assigned to the remaining variables. The variables corresponding to the leading 1's are called \emph{dependent} variables and the remaining variables are \emph{free}. Let $ \set{X_{F}} $ and $ \set{X} \setminus \set{X_{F}} $ denote the set of free and dependent variables respectively. Let $ \f = |\set{X_{F}}|$. Clearly $ \f = \numHashVars - rowrank(\mat{A}) $. For each possible assignment to the free variables we get an assignment to the dependent variables by propagating the values through the augmented matrix in $ \mathcal{O}(\numHashVars^{2}) $ time. Thus we can enumerate all $ 2^{\f} $ satisfying assignments to a system of linear equations $ \mat{A}.\vect{x} \voplus \vect{b}= \vect{y} $ if $ \mat{A} $ in \RRE form.

	 \subsection*{Gray Codes}
	 A Gray code~\cite{gray1953pulse} is an ordering of $ 2^{l} $ binary numbers for some $ l \ge 1 $ with the property that every pair of consecutive numbers in the sequence differ in exactly one bit. Thus starting from $ \vect{0}^{l} $ we can iteratively construct the entire Gray code sequence by flipping one bit in each step. We assume access to a procedure $ \nextGrayBit $ that in each call returns the position of the next bit that is to be flipped. Such a procedure can be implemented in constant time by a trivial modification of Algorithm L in \cite{knuth2004generating}.

\section{Related Work}
\label{sec:rel}
Propositional model counting has been of theoretical as well as practical interest over the years~\cite{Jerr,KarpLuby1983,Sang04combiningcomponent,Toda89}. Early investigations showed that both \#CNF and \#DNF are \#P-complete~\cite{Valiant79}. Consequently, approximation algorithms have been explored for both problems. A major breakthrough for approximate \#DNF was achieved by the seminal work of Karp and Luby~\cite{KarpLuby1983}, which provided a Monte Carlo-based FPRAS for \#DNF. The proposed FPRAS was improved by follow-up work of Karp, Luby and Madras~\cite{KarpLuby1989} and Dagum et al.~\cite{dagum2000optimal}, achieving the best known complexity of $\mathcal{O}(\numCubes\numVars\log(1/\delta)/\epsilon^{2})$. In this work, we bring certain ideas of Karp et al. into the hashing framework with significant adaptations.%
  
For \#CNF, early work on approximate counting resulted in hashing-based schemes that required polynomially many calls to an NP-oracle~\cite{Stockmeyer83,trevisan2002lecture}. No practical algorithms materialized from the these schemes due to the impracticality of the underlying NP queries. Subsequent attempts to circumvent hardness led to the development of several hashing and sampling-based approaches that achieved scalability but provided very weak or no guarantees~\cite{GSS06,gogate2007approximate}. Due to recent breakthroughs in the design of hashing-based techniques, several tools have been developed recently that can handle formulas involving hundreds of thousands of variables while providing rigorous formal guarantees. Overall, these tools can be broadly classified by their underlying hashing-based technique as: (i) obtain a constant factor approximation and then use identical copies of the input formula to obtain $\varepsilon$ approximations~\cite{EGSS13c}, or (ii) directly obtain $\varepsilon$ guarantees\cite{CMV13b,CMV16}. The first technique when applied to DNF formulas is not an FPRAS. In contrast, Chakraborty, Meel and Vardi~\cite{CMV16} recently showed that tools based on the latter approach, such as {\ScalApproxMC}, do provide FPRAS for \#DNF counting. Chakraborty et al. did not analyze the complexity of the algorithm in their work. We now provide a precise complexity analysis of {\ScalApproxMC} for \#DNF. To that end, we first describe the {\ApproxMC} framework on which $ \ScalApproxMC $ is built. 

	\subsection{ApproxMC Framework}
Chakraborty et al. introduced in~\cite{CMV13b} a hashing-based framework called {\ApproxMC} that requires linear (in $\numVars$) number of {\SAT} calls. Subsequently in {\ScalApproxMC}, the number of {\SAT} calls was reduced from linear to logarithmic (in $\numVars$). The core idea of {\ApproxMC} is to employ $2-$universal hash functions to partition the solution space into {\em roughly equal small} cells, wherein a cell is called {\em small} if it has less than or equal to {\hiThresh} solutions, such that {\hiThresh} is a function of $ \epsilon $. A {\SAT} solver is employed to check if a cell is small by enumerating solutions one-by-one until either there are no more solutions or we have already enumerated $\hiThresh+1$ solutions. Following the terminology of~\cite{CMV13b}, we refer to the above described procedure as {\BoundedSAT} (bounded SAT).  To determine the number of cells, {\ApproxMC} performs a search that requires $\mathcal{O}(\log \numVars)$ steps and the estimate is returned as the count of the solutions in a randomly picked small cell scaled by the total number of cells. To amplify confidence to the desired levels of $1-\delta$, {\ApproxMC} invokes the estimation routine $\mathcal{O}(\log \frac{1}{\delta})$ times and reports the median of all such estimates. Hence, the number of {\BoundedSAT} invocations is $\mathcal{O}(\log \numVars \log (\frac{1}{\delta}))$. 
	
\subsubsection*{FPRAS for \#DNF}
	The key insight of Chakraborty et al.~\cite{CMV16} is that the {\BoundedSAT} procedure can be done in polynomial time when the input formula to {\ApproxMC} is in DNF. In particular, the input to every invocation of {\BoundedSAT} is a formula that is a conjunction of the input DNF formula and a set of XOR constraints derived from the hash function. Chakraborty et al. observed that one can iterate over all the cubes of the input formula, substitute each cube into the set of XOR constraints separately, and employ Gaussian Elimination to enumerate the solutions of the simplified XOR constraints. Note that at no step would one have to enumerate more than {\hiThresh} solutions. Since Gaussian Elimination is a polynomial-time procedure, {\BoundedSAT} can be accomplished in polynomial time as well. Chakraborty et al. did not provide a precise complexity analysis of {\BoundedSAT}. We now provide such an analysis. We defer all proofs to the appendix. The following lemma states the time complexity of the {\BoundedSAT} routine. 
	
	\begin{lemma}
		\label{lem:bsat_time}
		The complexity of $ \BoundedSAT $ when the input formula to $ \ScalApproxMC $ is in DNF is $ \mathcal{O}(\numCubes\numVars^{3}+\numCubes\numVars^{2}/\epsilon^{2}) $.
		\qed
	\end{lemma}
 We can now complete the complexity analysis:
	\begin{lemma}
		\label{lem:dnfApprox_time}
		The complexity of $ \ScalApproxMC $ is $ \mathcal{O}((\numCubes\numVars^{3}+\numCubes\numVars^{2}/\epsilon^{2})\log \numVars\log(1/\delta)) $ when the input formula is in DNF.
		\qed
	\end{lemma}

\section{Efficient Hashing-based DNF Counter}
\label{sec:body}
We now present three key novel algorithmic innovations that allow us to design hashing-based FPRAS for \#DNF with complexity similar to Monte Carlo-based state-of-the-art techniques. We first introduce a new family of 2-universal hash functions that allow us to circumvent the need for expensive Gaussian Elimination. We then discuss the concept of Symbolic Hashing, which allows us to design hash functions over a space different than the assignment space, allowing us to achieve significant reduction in the complexity of search procedure for the number of the cells. Finally, we show that {\BoundedSAT} can be replaced by an efficient stochastic estimator. These three techniques allow us to achieve significant reduction in the complexity of hashing-based DNF counter without loss of theoretical guarantees. 

\subsection{Row-Echelon XOR Hash Functions}	\label{sec:rex}

The complexity analysis presented in Section~\ref{sec:rel} shows that the expensive Gaussian Elimination contributes significantly to poor time complexity of {\ScalApproxMC}. Since the need for Gaussian Elimination originates from the usage of $H_{XOR}$, we seek a family of 2-universal hash functions that circumvents this need. We now introduce a Row-Echelon XOR family of hash functions defined as $H_{REX}(\numHashVars,\numHashConstraints) = \{\mat{A}.\vect{x}\voplus\vect{b} \mid  \mat[A]{\numHashConstraints\times \numHashVars} = [\mat[I]{\numHashConstraints\times \numHashConstraints}\,\, :\,\, \mat[\D]{\numHashConstraints\times (\numHashVars-\numHashConstraints)}]\} $  where $ \mat{I} $ is the identity matrix, $ \mat{\D} $ and $\vect{b}$ are random 0/1 matrix and vector respectively. In particular,  we ensure that for every $ \mat{\D}[i,j] $ and $ \vect{b}[i] $ we have $ \Pr[\mat{\D}[i,j]=1] = \Pr[\mat{\D}[i,j]=0] = 0.5 $ and also $ \Pr[\vect{b}[i]=1] = \Pr[\vect{b}[i]=0] = 0.5 $. Note that $ \mat{\D} $ and $ \vect{b} $ completely define a hash function from $ H_{REX} $. The following theorem establishes the desired properties of universality for $H_{REX}$. The proof is deferred to Appendix. 

		\begin{theorem}
			\label{thm:rex_uni}
			$H_{REX}$ is 2-universal.
			\qed
		\end{theorem}

		\begin{algorithm}
			\caption{$\enumNext(\mat{\D},\vect{u},\vect{v},k)$}
			\label{alg:EREX}
			\begin{algorithmic}[1]
				\State $ \vect{v}' \gets \vect{v} $;
				\State $\vect{v}'[k] \gets \neg \vect{v}[k]$;
				\State $ \vect{u}' \gets \vect{u} \voplus \mat{\D}[.,k] $;
				\State \Return $ (\vect{u}',\vect{v}') $ 
			\end{algorithmic}
		\end{algorithm}
		The naive way of enumerating satisfying assignments for a given $ \mat[\D]{\numHashConstraints\times(\numHashVars-\numHashConstraints)}$, $ \vect[b]{\numHashConstraints} $, and $ \vect[y]{\numHashConstraints} $ is to iterate over all $2^{\f}$ assignments to the free variables in sequence starting from $\vect[0]{\f} $ to $ \vect[1]{\f}, $ where $ \f = (\numHashVars-\numHashConstraints) $. For each assignment $ \vect[v]{\f} $ to the free variables, the corresponding assignment to the dependent variables $ \vect[u]{\numHashVars-\f} $ can be calculated as $ \vect{u} = (\mat{\D} . \vect{v}) \voplus \vect{b} \voplus \vect{y} $, which requires $ O(\numHashConstraints\numHashVars$) time. Can we do better?
		
		We answer the above question positively by iterating over the $ 2^{\f} $ assignments to the free variables out of sequence. In particular, we iterate using the Gray code sequence for $ \f $ bits. The procedure is outlined in $ \enumNext $ (Algorithm \ref{alg:EREX}). The algorithm takes the hash matrix $ \mat{\D} $, an assignment to the free variables $ \vect{v} $, and an assignment to the dependent variables $ \vect{u} $ as inputs, and outputs the next free-variable assignment $ \vect{v}' $ in the Gray sequence and the corresponding assignment $ \vect{u}' $ to the dependent variables. $ k $ represents the position of the bit that is changed between $ \vect{v} $ and $ \vect{v}' $.		
		Thus $ \enumNext $ constructs a satisfying assignment to a Row-Echelon XOR hash function in each invocation in $ \mathcal{O}(\numHashVars) $ time.
	\subsection{Symbolic Hashing}
	\label{sec:symb}

	For DNF formulas, $\rf$ can be exponentially sparse compared to $\set{U}$, which is undesirable\footnote{Number of steps of $ \ScalApproxMC $ search procedure increases with sparsity}. 
	It is possible, however, to transform $ \set{U} $ to another space $ \set{U'} $ and the solution space $ \rf $ to $ \rf' $ such that the ratio $ |\set{U'}|/|\rf'| $ is polynomially bounded and $ |\rf| = |\rf'| $. For DNF formulas, the new universe $ \set{U'} $ is defined as $\set{U'}=\{(\vect{x}, \cube{i})\,\,|\,\, \vect{x} \models \cube{i}\}$. Thus, corresponding to each $\vect{x} \models \phi$ that satisfies cubes $\cube{i_{1}},..\cube{i_{\vect{x}}}$ in $ \phi $, we have the states $\{(\vect{x},\cube{i_{1}}),(\vect{x},\cube{i_{2}})..(\vect{x},\cube{i_{\vect{x}}})\}$ in $\set{U'}$.  Next, the solution space is defined as $\rf' = \{(\vect{x}, \cube{i})\,\, |\,\, \vect{x} \models \cube{i}\,\, \text{and} \,\, \forall j<i,  \vect{x} \not \models \cube{j}\}$ for a fixed ordering of the cubes. The definition of $\rf'$ ensures that $|\rf| = |\rf'|$.  This transformation is due to Karp and Luby \cite{KarpLuby1983}. 
	
	The key idea of Symbolic Hashing is to perform 2-universal hashing symbolically over the transformed space. In particular, the sampled hash function partitions the space $\set{U'}$ instead of $ \set{U} $. Therefore, we employ hash functions from $H_{REX}(\numHashVars,\numHashConstraints)$ over $\numHashVars = \numVars - \minClsWidth + \log \numCubes$ variables instead of $\numVars$ variables. Note that the variables of a satisfying assignment $\ha \in \{0,1\}^{\numHashVars}$ to the hash function are now different from the variables to a satisfying assignment $ \vect{x} \in \{0,1\}^{\numVars}$ of the input formula $ \phi $. We interpret $\ha$ as follows: the last $\log \numCubes$ bits of $ \ha $ are converted to a number $ i $ such that $ 1 \le i \le \numCubes $. Now $\cube{i}$ corresponds to a partial assignment of $\width{i}$ variables in that cube. For simplicity, we assume that each cube is of the same width $ \minClsWidth $.\footnote{We can handle non-uniform width cubes by sampling $ \cube{i} $ with probability $ \frac{2^{\numVars-\width{i}}}{\sum_{j=1}^{\numCubes}2^{\numVars-\width{j}}} $ instead of uniformly} The remaining $\numVars - \minClsWidth$  bits of $\ha$ are interpreted to be the assignment to the $\numVars-\minClsWidth$ variables not in $\cube{i}$ giving a complete assignment $\vect{x}$. Thus we get a pair $(\vect{x},\cube{i})$ from $\ha$ such that $ \vect{x} \models \cube{i} $. For a fixed ordering of variables and cubes we see that there is a bijection between $(\vect{x},\cube{i})$ and $\ha$ and hence the 2-universality guarantee holds over the partitioned space of $\set{U'}$.
	
	\subsection{Stochastic Cell-Counting}
	\label{sec:stoch}
	
	To estimate the number of solutions in a cell, we need to check for every tuple $(\vect{x},\cube{i})$ generated using symbolic hash function as described above: if $(\vect{x},\cube{i})  \in \rf'$. Such a check would require iteration over cubes $ \cube{j} $ for $ 1\le j\le(i-1) $ and returning no if $ \vect{x} \models \cube{j} $ for some $ j $ and yes otherwise. This would result in procedure with $ O(\numCubes\numVars) $ complexity. 
	
	Our key observation is that a precise count of the number of solutions in a cell is not required and therefore, one can employ a stochastic estimator for the number of solutions in a cell. We proceed as follows: we define the \emph{coverage} of an assignment $ \vect{x} $ as $ cov(\vect{x}) = \{ j | \vect{x} \models \cube{j} \} $. Note that $ \sum_{(\vect{x},\cube{i})\in \set{U}'} \frac{1}{|cov(\vect{x})|} = |\rf|$. 

		We define a random variable $ \cs $ as the number of steps taken to uniformly and independently sample from $ \{1,2,\ldots,\numCubes\} $, a number $ j $ such that $ \vect{x} \models \cube{j} $. For a randomly chosen $ j $, the probability $ \Pr[\vect{x}\models \cube{j}] = |cov(\vect{x})|/\numCubes$, which follows the Bernoulli distribution. The random variable $ \cs $ is the number of Bernoulli trials for the first success, which follows the geometric distribution. Therefore, $ \expect[\cs] = \numCubes/|cov(\vect{x})| $, and  $ \expect[\cs/\numCubes] = 1/|cov(\vect{x})|$. The estimator $ \cs/\numCubes $ has been previously employed by Karp et al.~\cite{KarpLuby1989}.  Here, we show that it can also be used for Stochastic Cell-Counting: we define the estimator for the number of solutions in a cell as $ \rfha = \sum_{(\vect{x},\cube{i})\in h^{-1}(\vect{y})} \cs/\numCubes $.%

		\begin{algorithm}
			\caption{$\SymbolicDNFApproxMCCore(\phi,\hiThresh)$}
			\label{alg:scalmccore}
			\begin{algorithmic}[1]
				\label{line:sfc-choose-h}
				\State $\minClsWidth \gets $ width of cubes;
				\State $ \numHashVars \gets \numVars - \minClsWidth + \log \numCubes$;
				\State $ \sI \gets \numVars - \minClsWidth - \log \hiThresh $;
				\State $ (\mat[\hat{\D}]{(\numHashVars-1)\times(\numHashVars-\sI)},\vect{\hat{b}} $,$ \vect{\hat{y}}) \gets \SampleBase(\numHashVars,\sI)$; 
				\State $\numHashConstraints \gets {\FibBinSearch}(\phi, \mat{\hat{\D}},\vect{\hat{b}}, \vect{\hat{y}}, \hiThresh,\sI, \numHashVars-1)$;
								
				\State $\solCount \gets {\BoundedSAT}(\phi, \mat{\hat{\D}},\vect{\hat{b}},\vect{\hat{y}}, \hiThresh, \numHashConstraints,\numHashVars, \sI)$;
				\State \Return $(2^{\numHashConstraints}, \solCount)$
			\end{algorithmic}
		\end{algorithm}
			\begin{algorithm}
				\caption{$\SymbolicDNFApproxMC(\phi,\varepsilon,\delta)$}
				\label{alg:scalmc}
				\begin{algorithmic}[1]
					\State $\hiThresh \gets 2*(1+ 9.84\left(1 + \frac{\varepsilon}{1+\varepsilon}\right)\left(1 + \frac{1}{\varepsilon}\right)^2)$; 
					\State $t \leftarrow  \lceil17\log_2 (3/\delta)\rceil$;
					\State $\text{EstimateList} \gets \emptyList$; $\iter \gets 0$;
					\Repeat
					\State $\iter \gets \iter+1$;
					\State {\footnotesize $\!(\cellCount, \solCount)\!\gets\! \SymbolicDNFApproxMCCore(\phi,\hiThresh)$;}
					\hspace{-3 em}\If {($\cellCount \ne \bot$)} $\AddToList(\text{EstimateList},\solCount \times \cellCount)$;
					\EndIf
					\Until {($\iter \ge t$)};
					\State $\mathrm{finalEstimate} \gets \FindMedian(\text{EstimateList})$;
					\State \Return $\mathrm{finalEstimate}$
				\end{algorithmic}
			\end{algorithm}
				
		\begin{algorithm}
			\caption{$ \SampleBase(\numHashVars,\sI) $}
			\label{alg:sb}
			\begin{algorithmic}[1]
				\State Sample $ \mat{\G} $ uniformly from $ \{0,1\}^{[\sI\times(\numHashVars-\sI)]} $;
				\State Sample uniformly an upper triangular matrix $ \mat[\M]{(\numHashVars-\sI-1)\times(\numHashVars-\sI)} $ with $ \mat{\M}[i,i] = 1 $ for all $ i $.
				\State $ \mat{\hat{\D}} \gets [\frac{\mat{\G}}{\mat{\M}}] $;
				\State Sample $ \vect{\hat{b}} $ and  $ \vect{\hat{y}} $ uniformly from $ \{0,1\}^{\numHashVars-1} $;
				\State \Return $ \mat{\hat{\D}},\vect{\hat{b}},\vect{\hat{y}}$
			\end{algorithmic}
		\end{algorithm}
		
		\begin{algorithm}
			\caption{$\FibBinSearch(\phi,\mat{\hat{\D}},\vect{\hat{b}},\vect{\hat{y}},\hiThresh, \sL, \sH)$}
			\label{alg:fib}
			\begin{algorithmic}[1]
				\State $\lowerFib \gets 0$; $\upperFib \gets \sH-\sL+1$; $ \numHashConstraints \gets \sL $;
				\State $\FailRecord[0] \gets 1$; $\FailRecord[\sH-\sL+1] \gets 0$;
				\State $\FailRecord[i] \gets \bot$ for all $i$ other than $0$ and $\sH-\sL+1$;
				\While{true}
				\State $C_{BSAT} \gets \BoundedSAT(\phi,\mat{\hat{\D}},\vect{\hat{b}},\vect{\hat{y}}, \hiThresh, \numHashConstraints,\numHashVars, \sI)$;
				\If {($C_{BSAT} \ge \hiThresh$)}
				\If {($\FailRecord[\numHashConstraints+1-\sL+1] = 0$)} \Return $\numHashConstraints+1$;
				\EndIf
				\State $\FailRecord[i] \gets 1$ for all $i \in \{1, \ldots \numHashConstraints-\sL+1\}$;
				\State  $\lowerFib \gets \numHashConstraints-\sL+1$;
				\State $\numHashConstraints \gets (\upperFib+\lowerFib)/2$;

				\Else 
				\If {($\FailRecord[\numHashConstraints-1-\sL+1] = 1$)} \Return $\numHashConstraints$;
				\EndIf
				\State $\FailRecord[i] \gets 0$ for all $i \in \{\numHashConstraints, \ldots \sH-\sL+1\}$;
				\State $\upperFib \gets \numHashConstraints-\sL+1$;
				\State $\numHashConstraints \gets (\upperFib+\lowerFib)/2 $;
				
				\EndIf
				\EndWhile
			\end{algorithmic}
		\end{algorithm}
	
		\subsection{The Full Algorithm}
		We now incorporate the above techniques into {\ScalApproxMC} and call the revised algorithm {\SymbolicDNFApproxMC}, which is presented as Algorithm~\ref{alg:scalmc}. First, note that expression for $ \hiThresh $ is twice that for $ \ScalApproxMC $. Then, in line 4, a matrix $ \mat{\hat{\D}}$ and vectors $\vect{\hat{b}}$ and $\vect{\hat{y}} $ are obtained, which are employed to construct an appropriate hash function and cell during the search procedure of  $ \SymbolicDNFApproxMCCore $.  $\SymbolicDNFApproxMC $ makes $ t = O(\log(1/\delta)) $ calls to $ \SymbolicDNFApproxMCCore $ (line 4-8) and returns median of all the estimates (lines 9-10) to boost the probability of success to $ 1-\delta $ .

		We now discuss the subroutine 	
		$ \SymbolicDNFApproxMCCore $, which is an adaptation of $ \ScalApproxMCCore $ but with significant differences. First, for DNF formulas with cube width $ w $, the number of solutions is lower bounded by $ 2^{\numVars-w} $. Therefore, instead of starting with $ 1 $ hash constraint, we can safely start with $ \sI = \numVars - w - \log \hiThresh $ constraints (lines 3-4). Thereafter, $ \SymbolicDNFApproxMCCore $ calls $ \FibBinSearch $ in line 5 to find the right number $ \numHashConstraints $ of constraints. The cell count with $ \numHashConstraints $ constraints is calculated in line 6 and the estimate $ (2^{\numHashConstraints}, \solCount) $ is returned in line 7.
		
		$ \SampleBase $ algorithm constructs the base matrix $ \mat{\hat{D}} $ and base vectors $\vect{\hat{b}}$ and $\vect{\hat{y}}$ required for sampling from $ H_{REX} $ family. $ \mat{\G} $ is a random matrix of dimension $ \sI\times(\numHashVars-\sI) $ and $ \mat{\M} $ is a random upper triangular matrix of dimension $ (\numHashVars-\sI-1)\times(\numHashVars-\sI) $ with all diagonal elements $ 1 $. In line 3, $ \mat{\hat{D}} $ is constructed as the vertical concatenation $  [\frac{\mat{\G}}{\mat{\M}}] $.
		
		$ \FibBinSearch $ (algorithm \ref{alg:fib}) performs a binary search to find the number of constraints $ \numHashConstraints $ at which the cell count falls below $ \hiThresh $. For DNF formula with cube width of $w$, since the number of solutions is bounded between $ 2^{\numVars-w} $ and $ \numCubes*2^{\numVars-w} $, we need to perform search for $\numHashConstraints$  between $\numVars-w$ and $\numVars-w+\log \numCubes$. Therefore, binary search can take at most $ O(\log\log\numCubes) $ steps to find correct $ \numHashConstraints $. 

	\begin{algorithm}
		\caption{$\BoundedSAT(\phi,\mat{\hat{\D}},\vect{\hat{b}}, \vect{\hat{y}},\hiThresh,\numHashConstraints, \numHashVars, \sI)$}
		\label{alg:BSAT}
		\begin{algorithmic}[1]
			\State $ count \gets 0 $;
			\State $ \mat{\D},\vect{b},\vect{y} \gets \Reduce(\mat{\hat{\D}},\vect{\hat{b}},\vect{\hat{y}},\numHashConstraints, \numHashVars, \sI)$; 

			\State $\vect{u}^{\numHashConstraints} \gets \vect{b}\voplus\vect{y}$;
			\State $ \vect{v}^{\numHashVars-\numHashConstraints} \gets \vect{0}^{\numHashVars-\numHashConstraints} $;
			\For{$(j = 0;\,\,j < 2^{\numHashVars-\numHashConstraints}; j++)$}
			\State $ \ha \gets [\vect{u} \,\,:\,\, \vect{v}] $;
			\State $ (\vect{x}, \cube{i}) = interpret(\ha) $;
			\State $ count = count + \CheckSAT(\vect{x}, \cube{i}, count, \hiThresh)$;
			\If {$ count \ge \hiThresh $}
			\Return $hiThresh$;
			\EndIf
			\State $ k \gets \nextGrayBit(\numHashVars-\numHashConstraints, j) $;
			\State $ (\vect{u},\vect{v}) \gets \enumNext(\mat{\D},\vect{u},\vect{v},k) $;
			\EndFor
			\State \Return $ count $
		\end{algorithmic}
	\end{algorithm}
		
	Symbolic Hashing is implemented in Algorithm \ref{alg:BSAT} ($ \BoundedSAT $). In line 2, we obtain a hash function from $H_{REX}(\numHashVars,\numHashConstraints)$ over $\numHashVars = \numVars - \minClsWidth + \log \numCubes$ variables by calling $ \Reduce $. We assume access to a procedure $ \nextGrayBit $ in line 10 that returns the position of the bit that is flipped between two consecutive assignments. A satisfying assignment $ \ha $ to the hash function is constructed in line 6. $ \ha $ is interpreted to generate a pair $ (\vect{x}, \cube{i}) $ in line 7 which is checked for satisfiability in line 8. The final cell count is returned in line 12.
	
	\begin{algorithm}
		\caption{$\CheckSAT(\vect{x}, \cube{i}, count, \hiThresh)$}
		\label{alg:CSAT2}
		\begin{algorithmic}[1]
			\State $ \cs \gets 0 $;
			\While{$count+\cs/m < \hiThresh$}
			\State Uniformly sample $j$ from  $ \{1,2,..,\numCubes\} $;
			\State $ \cs \gets \cs+1 $;
			\If {$ \vect{x} \models \cube{j} $}
			\State \Return $\cs / \numCubes$;
			\EndIf
			\EndWhile
			\State \Return $\cs / \numCubes$
		\end{algorithmic}
	\end{algorithm}
	
	In $ \CheckSAT $ (algorithm \ref{alg:CSAT2}), we implement the stochastic cell counting procedure. The key idea is to  sample cubes uniformly at random from $ \{1,2.\dots \numCubes\} $ till a cube $ \cube{j} $ is found such that $ \vect{x} \models \cube{j} $ (lines 2-5). The number of cubes sampled $ \cs $ divided by total number of cubes $ \numCubes $ is the estimate returned (line 6).
	\label{sec:imp}
	\begin{algorithm}
		\caption{$\Reduce(\mat{\hat{\D}},\vect{\hat{b}},\vect{\hat{y}},\numHashConstraints, \numHashVars, \sI) $}
		\label{alg:red}
		\begin{algorithmic}[1]
			\State $ \mat{\hat{\D}'} \gets \mat{\hat{\D}}[0:\numHashConstraints,0:\numHashVars-\sI] $; $ \vect{b} \gets \vect{\hat{b}}[0:\numHashConstraints] $;
			\State $ \vect{y} \gets \vect{\hat{y}}[0:\numHashConstraints] $;
			\For {$ (i = \sI;\,\,i < \numHashConstraints; i++) $}
			\For {$ (j = 0;\,\,j < i; j++) $}
			\If {$ \mat{\hat{\D}'}[j,(i-\sI)] == 1 $}
			\State $ \mat{\hat{\D}'}[j,.] \gets \mat{\hat{\D}'}[j,.] \voplus \mat{\hat{\D}'}[i,.] $; 
			\State $ \vect{b}[j] \gets \vect{b}[j] \oplus \vect{b}[i] $;
			\State $ \vect{y}[j] \gets \vect{y}[j] \oplus \vect{y}[i] $;
			\EndIf
			\EndFor
			\EndFor
			\State $ \mat{\D} \gets \mat{\hat{\D}'}[0:\numHashConstraints,\numHashConstraints-\sI:\numHashVars-\sI] $;
			\State \Return $ \mat{\D}, \vect{b}, \vect{y} $
		\end{algorithmic}
	\end{algorithm}
	Procedure $ \FibBinSearch $ in $ \SymbolicDNFApproxMC $ is based upon the $ \FibBinSearch $ in $ \ScalApproxMC $\cite{CMV16}. As noted in the analysis of {\ScalApproxMC}, 
	such a logarithmic search procedure requires that  the solution space for a hash function with $ \numHashConstraints + 1$ hash constraints is a subset of the solution space with $ \numHashConstraints$ hash constraints. 
	Furthermore, we want to preserve Row-Echelon nature of the resulting hash constraints. To this end, we first construct $\mat[\D]{\numHashVars\times(\numHashVars-\numHashConstraints)}$ and $ \vect[b]{\numHashVars-1} $ as follows:
	
	To seed the construction procedure, in $ \SampleBase $ (algorithm \ref{alg:sb}) we first randomly sample a 0/1 vector $\vect{\hat{b}}$ of size $ \numHashVars-1 $ which is the maximum number of hash constraints possible. We then construct a 0/1 matrix $\mat{\hat{\D}}$ as follows: $\mat[\hat{\D}]{(\numHashVars-1)\times(\numHashVars-\sI)} = [\frac{\mat{\G}}{\mat{\M}}]$ where matrix $ \mat[\G]{\sI\times(\numHashVars-\sI)} $ is a random 0/1 matrix with $ \sI $ rows, and matrix $ \mat[\M]{(\numHashVars-\sI)\times(\numHashVars-\sI)} $ is defined as  as follows:
	\begin{itemize}
		\item $ \mat{\M}[i,j] = 1 $ if $ i=j $
		\item $ \mat{\M}[i,j] = 0 $ if $ i>j $
		\item $ \Pr[\mat{\M}[i,j] = 1] = \Pr[\mat{\M}[i,j] = 0] = 0.5  $ if $ i<j $
	\end{itemize}
		
	The reason for this definition of $ \mat{\hat{\D}} $ is that for DNF counting we have a good lower bound on the number of hash constraints we can start with. The number of rows in $ \mat{\G} $ corresponds to this lower bound. The definition of $ \mat{\M} $ ensures that the rows of $ \mat{\M} $ are linearly independent which results in a monotonically shrinking solution space.
	
	The $ \Reduce $ procedure (algorithm \ref{alg:red}) takes $ \mat{\hat{\D}} $,$ \vect{\hat{b}} $ and $ \vect{\hat{y}} $ and a number $ \numHashConstraints $ as input and returns $ \mat{\D} $,$\vect{b}$ and cell $ \vect{y} $ such that $ (\mat{\D},\vect{b}) $ represents a hash function from $ H_{REX} $ with $ \numHashConstraints $ constraints and $ \vect{y} $ represents a cell. A precondition for $ \Reduce $ is $ \sI \le \numHashConstraints \le \numHashVars-1$. In lines 1 and 2, the first $ \numHashConstraints $ rows of $ \mat{\hat{\D}} $ and first $ \numHashConstraints $ elements of $ \vect{\hat{b}} $ and $ \vect{\hat{y}} $ are selected as $ \mat{\hat{\D}'} $, $ \vect{b} $ and $ \vect{y} $ respectively. The first $ \sI $ rows of $ \mat{\hat{\D}'} $ form the matrix $ \mat{\G} $ in the definition of $ \mat{\hat{\D}} $ and the remaining $ \numHashConstraints - \sI $ rows of $ \mat{\hat{\D}'} $ are the first $ \numHashConstraints - \sI $ rows of matrix $ \mat{\M} $. Each row from $ \sI $ to $ \numHashConstraints $ is used to reduce the preceding rows in lines 5 to 8 so that the only non-zero elements of the first $ \numHashConstraints - \sI $ columns are the leading 1s in rows $ \sI $ to $ \numHashConstraints $. Thus $ \Reduce $ ensures that for a given $ \mat{\hat{\D}} $,$ \vect{b} $ and $ \vect{\hat{y}} $, the solution space of $ \mat[\D]{\numHashConstraints\times(\numHashVars-\numHashConstraints)} $,$ \vect[b]{\numHashConstraints} $ and $ \vect[y]{\numHashConstraints} $ is a superset of solution space of $ \mat[\D]{(\numHashConstraints+1)\times(\numHashVars-\numHashConstraints-1)} $,$ \vect[b]{\numHashConstraints+1} $ and $ \vect[y]{\numHashConstraints+1} $ for all $ \numHashConstraints $.
\section{Analysis}
\label{sec:ana}
	In order to prove the correctness of $ \SymbolicDNFApproxMC $, we first state the following helper lemma. We defer the proofs to the Appendix.
	
	\begin{lemma}
		\label{lem:stoc1}
		For every $ 1 \le \numHashConstraints \le \numHashVars $ and let $ \mu_{\numHashConstraints} = |\rf|/2^{\numHashConstraints} $. For every $ \beta > 0 $ and $ 0 < \epsilon < 1 $ we have 
		\begin{enumerate}
			\item $ \Pr[|\rfha - \mu_{\numHashConstraints}| > \frac{\epsilon}{(1+\epsilon)}\mu_{\numHashConstraints}] \le \frac{2}{\frac{\epsilon^{2}}{(1+\epsilon)^{2}}\mu_{\numHashConstraints}} $
			\item $ \Pr[\rfha\le \beta\mu_{\numHashConstraints}] \le \frac{2}{2+(1-\beta^{2})\mu_{\numHashConstraints}}$
		\end{enumerate}
		\qed
	\end{lemma}
	The difference in lemma \ref{lem:stoc1} and lemma 1 in \cite{CMV16} is that the probability bounds differ by a factor of 2. We account for this difference by making $ \hiThresh $ in $ \SymbolicDNFApproxMC $ twice the value of $ \hiThresh $ in $ \ScalApproxMC $. Therefore the rest of the proof of Theorem \ref{thm:stoc_correct} (below) is exactly the same as the proof of Theorem 4 of \cite{CMV16}. For completeness, we first restate lemmas 2 and 3 from \cite{CMV16} below. 
			
			In the following, $ T_{\numHashConstraints} $ denotes the event $ (\rfha < \hiThresh) $, and $ L_{\numHashConstraints} $ and $ U_{\numHashConstraints} $ denote the events $ (\rfha < \frac{|\rf|}{(1+\epsilon)2^{\numHashConstraints}}) $ and $ (\rfha > \frac{|\rf|}{2^{\numHashConstraints}}(1+\frac{\epsilon}{1+\epsilon})) $ respectively. $ \numHashConstraints^{*} $ denotes the integer $ \floor{\log_{2}|\rf| - \log_{2} (4.92(1+\frac{1}{\epsilon})^{2})} $
						
			\begin{lemma}
			\label{lem:stoc2}
				The following bounds hold:
				\begin{enumerate}
					\item $ \Pr[T_{\numHashConstraints^{*}-3}] \le \frac{1}{62.5}$
					\item $ \Pr[L_{\numHashConstraints^{*}-2}] \le \frac{1}{20.68} $
					\item $ \Pr[L_{\numHashConstraints^{*}-1}] \le \frac{1}{10.84}$
					\item $ \Pr[L_{\numHashConstraints^{*}}\cup U_{\numHashConstraints^{*}}] \le \frac{1}{4.92} $
				\end{enumerate}
				\qed
			\end{lemma}
			
			Let $ B $ denote the event that $ \SymbolicDNFApproxMC $ returns a pair $ (2^{\numHashConstraints},nSols) $ such that $ 2^{\numHashConstraints}*nSols $ does not lie in the interval $ [\frac{\set{|\rf|}}{1+\epsilon}, |\rf|(1+\epsilon) ]$.
			
			\begin{lemma}
			\label{lem:stoc3}
				$ \Pr[B] \le 0.36 $
				\qed
			\end{lemma}
		
		\begin{theorem}
			\label{thm:stoc_correct}
			Let $ \SymbolicDNFApproxMC(\phi, \epsilon, \delta) $ return count $ c $. Then $ \Pr[|\set{R_{\phi}}|/(1+\epsilon) \le c \le (1+\epsilon)|\set{R_{\phi}}|] \ge 1-\delta $.
			\qed
		\end{theorem}
				
		Theorem \ref{thm:stoc_correct} follows from lemmas \ref{lem:stoc1}, \ref{lem:stoc2} and \ref{lem:stoc3}	and noting that $ \SymbolicDNFApproxMC $ boosts the probability of correctness of the count returned by $ \SymbolicDNFApproxMCCore $ to $ 1-\delta $ by using median of $ t = O(\log (1/\delta) )$ calls.
		
		\begin{theorem}
			\label{thm:stoc_time}
			$ \SymbolicDNFApproxMC $ runs in $ \tilde{O}(\numCubes\numVars\log(1/\delta)/\epsilon^{2}) $ time.\footnote{We say $f(n) \in \tilde{O}(g(n))$ if $\exists k: f(n) \in \mathcal{O}(g(n) \log^k (g(n))$}
			\qed
		\end{theorem}
%

%
%
\section{Conclusion}\label{sec:conc}
Hashing-based techniques have emerged as a promising approach to obtain counting algorithms and tools that scale to large instances while providing strong theoretical guarantees. This has led to an interest in designing hashing-based algorithms for counting problems that are known to be amenable to fully polynomial randomized approximation schemes. The prior hashing-based approach~\cite{CMV16} provided FPRAS for DNF but with complexity much worse than state-of-the-art techniques. In this work, we introduced (i) Symbolic Hashing, (ii) Stochastic Cell-Counting, and (iii) a new 2-universal family of hash functions, and obtained a hashing-based FPRAS for \#DNF with complexity similar to state-of-the-art. %

Given the recent interest in hashing-based techniques and generality of our contributions, we believe concepts introduced in this paper can lead to design of hashing-based techniques for other classes of constraints. For example, all prior versions of $ \ApproxMC $ relied on deterministic SAT solvers for exactly counting the solutions in a cell for \#CNF. The technique of Stochastic Cell-Counting opens up the door for the usage of probabilistic SAT solvers for \#CNF.  Furthermore, a salient feature of the $ H_{REX} $ family is the sparsity of its hash functions. In fact, the sparsity increases with the addition of constraints. Sparse hash functions have been shown to be desirable for efficiently solving CNF+XOR constraints~\cite{IMMV15,EGSS14a,ghss07:shortxors}. An interesting direction for future work is to test $ H_{REX} $ family with CNF formulas.%

\subparagraph*{Acknowledgements}
The authors thank Jeffrey Dudek, Supratik Chakraborty and Dror Fried for valuable discussions. Work supported in part by NSF grants CCF-1319459 and IIS-1527668, by NSF Expeditions in Computing project "ExCAPE: Expeditions in Computer Augmented Program Engineering", and by an IBM Graduate Fellowship. Kuldeep S. Meel is supported by the IBM PhD Fellowship and the Lodieska Stockbridge Vaughn Fellowship.

\section*{Appendix}
	Proof of Lemma \ref{lem:bsat_time}
	\begin{proof}
		When the input formula $ \phi $ to $ \ScalApproxMC $ is in DNF, $ \BoundedSAT $ is invoked with a formula of the form $ \phi \land \psi $ where $ \psi $ is a conjunction of XOR constraints. For each cube $ \cube{i} $, $ \BoundedSAT $ proceeds by performing Gaussian Elimination on $ \cube{i} \land \psi$. Since the number of XOR constraints can be $ \mathcal{O}(\numVars) $, Gaussian elimination can take $\mathcal{O}(\numVars^{3}) $ time resulting in a cumulative complexity of $ \mathcal{O}(\numCubes\numVars^{3}) $ for all cubes. 
		
		At most  $ \hiThresh = \mathcal{O}(1/\epsilon^{2})$ solutions to each $ \cube{i} \land \psi$ may have to be enumerated and each enumeration requires $ \mathcal{O}(\numVars^{2}) $ time. Therefore the complexity of enumeration is $ \mathcal{O}(\numCubes\numVars^{2}/\epsilon^{2}) $.
		Thus the $ \BoundedSAT $ runs in $ \mathcal{O}(\numCubes\numVars^{3}+\numCubes\numVars^{2}/\epsilon^{2}) $ time in the worst case.
		\qed
	\end{proof}
	Proof of Lemma \ref{lem:dnfApprox_time}
	\begin{proof}
		Theorem 4 in \cite{CMV16} shows that $ \ScalApproxMC $ makes $ \mathcal{O}(\log \numVars\log(1/\delta)) $ calls to $ \BoundedSAT $. Substituting the complexity of $ \BoundedSAT $ from lemma \ref{lem:bsat_time}, we get $\mathcal{O}((\numCubes\numVars^{3}+\numCubes\numVars^{2}/\epsilon^{2})\log \numVars\log(1/\delta)) $ complexity for $ \ScalApproxMC $.
		\qed 
	\end{proof}
	Proof of Theorem \ref{thm:rex_uni}:
	\begin{proof}
		Let $ h $ be a random hash function from $H_{REX}(\numHashVars,\numHashConstraints) $ with $ \mat[A]{\numHashConstraints \times \numHashVars} $ as its matrix. 
		Let $ \vect{x}_{1}, \vect{x}_{2}$ be any two assignments such that $ \vect{x}_{1} \ne \vect{x}_{2} $. To prove 2-universality of $ H_{REX} $, we need to show that for all $ \vect{x} $ and $ \vect{y} $:
		\begin{eqnarray}
		\Pr[h(\vect{x}_{1})= h(\vect{x}_{2})] \le \frac{1}{2^{\numHashConstraints}}			\label{eqn:2-uni}\\
		\Pr[h(\vect{x}) = \vect{y}] = \frac{1}{2^{\numHashConstraints}}\label{eqn:uni}
		\end{eqnarray}
		
		Equation \ref{eqn:uni} follows from the random choice of $ \vect{b} $. In particular, for every $ \vect{x} $ and $ \vect{y} $ for a chosen $ \mat{A} $, there is a unique $ \vect{b} $ such that $ \mat{A}.\vect{x}\voplus\vect{b} = \vect{y} $. Since there are $ 2^{\numHashConstraints} $ possible choices for $ \vect{b} $, we have $ \Pr[\mat{A}.\vect{x}\voplus\vect{b} = \vect{y} ] = \frac{1}{2^{\numHashConstraints}} $.
		
		Let $\vect{x} = \vect{x}_{1} \voplus \vect{x}_{2}$. Note that $ \vect{x} \not = \vect[0]{\numHashVars} $ since $ \vect{x}_{1} \ne \vect{x}_{2} $. We prove that $ \Pr[h(\vect{x})= \vect{0}^{\numHashVars}] \le 1/2^{\numHashConstraints} $ which is equivalent to \ref{eqn:2-uni}. This is the same as showing $ \Pr[\mat{A}.\vect{x} = \vect[0]{\numHashVars}] \le 1/2^{\numHashConstraints} $. We can write $ \mat{A} = [\mat[I]{\numHashConstraints\times \numHashConstraints}\,\, :\,\, \mat[\D]{\numHashConstraints\times (\numHashVars-\numHashConstraints)}] $ and $ \vect{x} $ as $ \vect{x} = [\vect[u]{\numHashConstraints}\,\,:\,\, \vect[v]{(\numHashVars-\numHashConstraints)}] $ in block notation. Then we have $ \mat{A}.\vect{x} = \mat{I}.\vect{u} \voplus \mat{\D}.\vect{v} $. Since $ \vect{x} \ne \vect{0} $, either $ \vect{u} \ne \vect{0}$ or $ \vect{v} \ne \vect{0} $ leading to the following three cases:
		\begin{description}
			
			\item[Case 1:] If $ \vect{u} \not = \vect[0]{\numHashConstraints} $ and $ \vect{v} = \vect[0]{\numHashVars-\numHashConstraints} $, we get $ \mat{I}.\vect{u} \not = \vect[0]{\numHashConstraints} $ and $ \mat{\D}.\vect{v} = \vect[0]{\numHashConstraints} $. Therefore $ \mat{A}.\vect{x} = \mat{I}.\vect{u} \voplus \mat{\D}.\vect{v} \not = \vect[0]{\numHashConstraints} $.
			\item[Case 2:] If $ \vect{u} = \vect[0]{\numHashConstraints} $ and $ \vect{v} \not = \vect[0]{\numHashVars-\numHashConstraints} $, we get $ \mat{I}.\vect{u} = \vect[0]{\numHashConstraints} $. From the proof of Theorem 1 in \cite{markowsky1978analysis} we have $ \prob [\mat{\D}.\vect{v} = \vect[0]{\numHashConstraints}] = \frac{1}{2^{\numHashConstraints}} $. Therefore $ \prob [\mat{A}.\vect{x} = \vect[0]{\numHashConstraints}] = \frac{1}{2^{\numHashConstraints}} $
			\item[Case 3:] If $ \vect{u} \not = \vect[0]{\numHashConstraints} $ and $ \vect{v} \not = \vect[0]{\numHashVars-\numHashConstraints} $, we get $ \mat{I}.\vect{u} \not = \vect[0]{\numHashConstraints} $ and $ \prob [\mat{\D}.\vect{v} = \vect[0]{\numHashConstraints}] = \frac{1}{2^{\numHashConstraints}} $. Therefore $ \prob [\mat{A}.\vect{x} = \vect[0]{\numHashConstraints}] = \frac{1}{2^{\numHashConstraints}} $
		\end{description}
		\qed
	\end{proof}
	Proof of Lemma \ref{lem:stoc1}:
	\begin{proof}
	The \emph{coverage} of an assignment is $ cov(\vect{x}) = \{j | \vect{x} \models \cube{j}\} $. We have $ \Pr[\vect{x} \models \cube{j}] = |cov(\vect{x})| / \numCubes $ when $ j $ is drawn uniformly at random from $\{1,2.\dots\numCubes\}$. Also, $ \sum_{(\vect{x},\cube{i})\in \set{U'}} \frac{1}{|cov(\vect{x})|} = |\rf| $, and $ \expect[\cs] = \numCubes / |cov(\vect{x})|$ and $ \expect[\cs^{2}] = \frac{2\numCubes^{2}}{|cov(\vect{x})|^{2}} - \frac{\numCubes}{|cov(\vect{x})|} $.
	
	Let $ \gamma_{(\vect{x},\cube{i}),\vect{y}} $ be a random variable such that $ \gamma_{(\vect{x},\cube{i}),\vect{y}} = \cs/\numCubes $ if $ h(\vect{x}) = \vect{y} $ and $ (\vect{x},\cube{i}) = interpret(\vect{z}) $, and $ \gamma_{(\vect{x},\cube{i}),\vect{y}} = 0 $ otherwise, where the number of constraints in $ h $ is $ \numHashConstraints $. Let $ \eta = \Pr[h(\vect{x}) = \vect{y}] = 1/2^{\numHashConstraints}$. Then $ \expect[\gamma_{(\vect{x},\cube{i}),\vect{y}}] = \eta*\expect[\frac{c_{\vect{x}}}{\numCubes}] = \frac{\eta}{|cov(\vect{x})|} $.
	
	Now, $ \var[\gamma_{(\vect{x},\cube{i}),\vect{y}}] = \expect[\gamma_{(\vect{x},\cube{i}),\vect{y}}^{2}] - (\expect[\gamma_{(\vect{x},\cube{i}),\vect{y}}])^{2} $. But $\expect[\gamma_{(\vect{x},\cube{i}),\vect{y}}^{2}] = \eta*\expect[(\cs/\numCubes)^{2}] = \eta(\frac{2}{|cov(\vect{x})|^{2}}-\frac{1}{\numCubes*|cov(\vect{x})|})$ and $ (\expect[\gamma_{(\vect{x},\cube{i}),\vect{y}}])^{2} = \frac{\eta^{2}}{|cov(\vect{x})|^{2}} $. Substituting back we get $ \var[\gamma_{(\vect{x},\cube{i}),\vect{y}}] = \frac{2\eta-\eta^{2}}{|cov(\vect{x})|^{2}} - \frac{\eta}{\numCubes*|cov(\vect{x})|}$.
	
	Define $ \tau_{\vect{y}} = \sum_{(\vect{x},\cube{i})\in \set{U'}} \gamma_{(\vect{x},\cube{i}),\vect{y}} $. Clearly $ \tau_{\vect{y}} = |\rfha| $. We have  $ \expect[\tau_{\vect{y}}] = \sum_{(\vect{x},\cube{i})\in U'} \expect[\gamma_{(\vect{x},\cube{i}),\vect{y}}] = \sum_{(\vect{x},\cube{i})\in U'} \frac{\eta}{|cov(\vect{x})|} = \eta|\rf| $. Also since $\gamma_{(\vect{x},\cube{i}),\vect{y}}$s are 2-universal, $ \var[\tau_{\vect{y}}] \le \sum_{(\vect{x},\cube{i})\in U'} \var[\gamma_{(\vect{x},\cube{i}),\vect{y}}] = (2\eta-\eta^{2})\sum_{(\vect{x},\cube{i})\in U'} \frac{1}{|cov(\vect{x})|^{2}} - \frac{\eta}{\numCubes}\sum_{(\vect{x},\cube{i})\in U'} \frac{1}{|cov(\vect{x})|}$. But $ \sum_{(\vect{x},\cube{i})\in U'} \frac{1}{|cov(\vect{x})|^{2}} \le |\rf| $. Therefore $ \var[\tau_{\vect{y}}] \le 2\eta|\rf| $ which implies $ \var[\tau_{\vect{y}}] \le 2 \expect[\tau_{\vect{y}}]$. 
	
	Applying Chebyshev's inequality we get $ \Pr[|\rfha-\mu_{\numHashConstraints}|\ge \frac{\epsilon}{1+\epsilon}\mu_{\numHashConstraints}] \le \frac{\var[\tau_{\vect{y}}]}{\frac{\epsilon^{2}}{(1+\epsilon)^{2}}\mu_{\numHashConstraints}^{2}}$. Rearranging the terms and simplifying, we get the first part of the lemma.
	
	Using Paley-Zygmund inequality, we get $\Pr[\rfha\le \beta\mu_{\numHashConstraints}] \le 1-\frac{(1-\beta)^{2}\mu_{\numHashConstraints}^{2}}{\var[\tau_{\vect{y}}]+(1-\beta^{2})\mu_{\numHashConstraints}^{2}}$ from which we get the second part of the lemma.
	\qed
	\end{proof}
	Proof of Theorem \ref{thm:stoc_time}
	\begin{proof}
		$ \SymbolicDNFApproxMC $ makes $ \mathcal{O}(\log 1/\delta) $ calls to $ \SymbolicDNFApproxMCCore $. 
		
		$ \SymbolicDNFApproxMCCore $ samples a hash matrix $ \mat{\hat{\D}} $ in $ \mathcal{O}(\numVars (\log\numCubes + \log(1/\epsilon^{2}))) $ time and makes one call each to $ \FibBinSearch $ and $ \BoundedSAT $. $ \FibBinSearch $ search in turn makes up to $ \mathcal{O}(\log \log \numCubes) $ calls to $ \BoundedSAT $. $ \BoundedSAT $ first reduces the matrix $ \mat{\hat{\D}} $ to $ \mat{\D} $ in $ O(\numVars(\log\numCubes + \log(1/\epsilon^{2})^{2}) $ time by calling $ \Reduce $. Next, the check in line 9 of {\BoundedSAT} ensures that at most $m \times \hiThresh$ calls are made to the check in line 5 of {\CheckSAT} during one execution of {\BoundedSAT}. Every assignment is enumerated in $\mathcal{O}(\numVars)$ and the check in line 5 of {\CheckSAT} is performed in $O(\numVars)$ time. Therefore, 
		the time complexity of $ \BoundedSAT $ is  $ O(\numVars\times \numCubes \times \hiThresh) $. Hence, the time complexity of each invocation of $ \SymbolicDNFApproxMCCore $ is $ \tilde{O}(\numCubes\numVars/\epsilon^{2})$, which implies that the time complexity of $ \SymbolicDNFApproxMC $ is $\tilde{O}(\numCubes\numVars\log(1/\delta)/\epsilon^{2}) $.
		\qed
	\end{proof}

\end{document}